\documentclass[reprint,superscriptaddress,amsmath,amssymb,aps,twocolumn,floatfix]{revtex4-2}

\usepackage[section]{placeins}

\usepackage{amsmath,amssymb}
\usepackage{wasysym}
\usepackage{lipsum}
\usepackage{csquotes}
\usepackage{graphicx,xcolor}
\usepackage{dcolumn}
\usepackage{bm}

\usepackage{printlen} 
\usepackage{setspace}
\usepackage{hyperref}

\usepackage[capitalise]{cleveref}

\makeatletter
\AddToHook{cmd/appendix/before}{\def\cref@section@alias{appendix}\def\cref@subsection@alias{appendix}}
\makeatother

\usepackage{physics}
\usepackage{xspace}

\newcommand{\ignore}[1]{}

\usepackage{color}

\let\oldchi\chi
\let\chi\undefined
\DeclareRobustCommand{\chi}{{\mathpalette\irchi\relax}}
\newcommand{\irchi}[2]{\raisebox{\depth}{$#1\oldchi$}} 

\newcommand{\cs}{~}

\newcounter{myfigpanel}[figure]
\newcounter{myfigpanelonly}[figure]

\newcommand{\panelletter}[1]{\refstepcounter{myfigpanel}\label{#1}\refstepcounter{myfigpanelonly}\label{onlyletter:#1}\alph{myfigpanel}}
\newcommand{\panel}[1]{(\protect\panelletter{#1})}

\crefalias{myfigpanel}{figure}
\crefname{myfigpanelonly}{panel}{panels}

\let\origcaption\caption
\let\caption\undefined
\DeclareRobustCommand{\caption}[1]{\origcaption{\protect\setcounter{myfigpanel}{0}\protect\setcounter{myfigpanelonly}{0}#1}}

\newcommand{\MPIDS}{\affiliation{Max Planck Institute for Dynamics and Self-Organization, Göttingen, Germany}}

\newcommand{\RPCTP}{\affiliation{Rudolf Peierls Centre for Theoretical Physics, University of Oxford, Oxford OX1 3PU, United Kingdom}}

\begin{document}

\title{Orientational lineage memory and mechanical ordering during diffusion-limited growth}

\author{Ilias-Marios Sarris}%
\MPIDS

\author{Ramin Golestianian}
\MPIDS
\RPCTP

\author{Philip Bittihn}
\MPIDS
\email{philip.bittihn@ds.mpg.de}

\begin{abstract}
Growth and shape formation in crowded multicellular assemblies arise from the interplay of chemical gradients, single-cell expansion and mechanical interactions, making it essential to understand how these processes jointly shape collective organization. Using a particle-based model that resolves nutrient fields as well as cellular orientations and their inheritance, we investigate how orientational order emerges within expanding fronts whose morphology is set by nutrient limitation. We identify a transition in nematic order controlled by front morphology, with orientational memory influencing alignment only on one side of this transition. Under strong inheritance, orientational order varies non-monotonically: both thin active layers (fingering morphologies) and thick active layers (flat fronts) produce strong alignment, whereas intermediate cases are less ordered. Analysis of velocities, reorientation statistics, and stress anisotropies shows that this behavior reflects a shift from inheritance-dominated to mechanically driven alignment that overrides lineage memory. The resulting differences in front speed produce a fitness advantage of orientational memory only in the diffusion-limited, memory-dominated regime. These findings elucidate how nutrient supply, mechanical interactions, and single-cell expansion together shape self-organization during growth.
\end{abstract}
\maketitle

\section{Introduction}

Multicellular systems, from epithelial layers to bacterial colonies, exhibit rich self-organization driven by their inherently active nature\cs\cite{hallatschekProliferatingActiveMatter2023}. Among the various forms of cellular activity operating across scales, growth occupies a special role. Unlike other active processes, growth injects both energy and new biomass into the system, and recent efforts have turned toward understanding the role of growth in shaping large-scale organization and collective dynamics\cs\cite{gelimsonCollectiveDynamicsDividing2015a,Meacock2020,Welker2021,Zhou2022}. The resulting phenomena span a wide range, including interfacial patterns\cs\cite{yanMechanicalInstabilityInterfacial2019}, to nematic alignment and defect dynamics in dense assemblies\cs\cite{nijjerMechanicalForcesDrive2021,nijjerBiofilmsSelfshapingGrowing2023,guillamatIntegerTopologicalDefects2022}, and, when combined with motility coordinated chemotactic behaviors at tissue and colony scales\cs\cite{gelimsonCollectiveDynamicsDividing2015a} and mixing transitions\cs\cite{sunkelTangentialDiffusionMotilityinduced2024}.

Cell proliferation is regulated by a combination of spatially varying chemical\cs\cite{zhuMetabolicRegulationCell2019,wangShapeGrowingFront2017,Welker2021,stevanovicDynamicalModelAntibiotic2024b,stevanovicNutrientGradientsMediate2022} and mechanical cues\cs\cite{silveiraMechanicalInteractionsTissue2025,ladouxMechanobiologyCollectiveCell2017a}, encompassing a wide range of biochemical signals, environmental factors, and physical forces. These variations generate differential growth, which is known to produce a broad spectrum of morphologies including branching\cs\cite{fujikawaFractalGrowthBacillus1989,ranaSpreadingNonmotileBacteria2017,wangShapeGrowingFront2017}, buckling and folding\cs\cite{trejoElasticityWrinkledMorphology2013,tozluogluPlanarDifferentialGrowth2019}, and diverse invasive or collective behaviors relevant to wound healing, spreading monolayers, bacterial colony expansion, and tumor progression\cs\cite{petitjeanVelocityFieldsCollectively2010,Audoin2022}.

Beyond changes in growth rate, packing, shape, and orientation of cells within an expanding tissue or colony play a central role in determining its emergent patterns\cs\cite{isenseeSensitiveParticleShape2025}. In most biological systems, growth and division are not isotropic: cells add mass and divide along preferred axes, generating a microscopic polarity that directs forces and transmits stresses. This local mechanical polarity can couple strongly to collective dynamics and, in turn, be reshaped by them, creating a feedback loop between cellular orientation and large-scale pattern formation\cs\cite{lemkeDynamicChangesEpithelial2021,kimCoordinationCellPolarity2018,gorelovaPlantCellPolarity2021}. 

The origin of such orientational bias, however, can vary widely across systems. In dense bacterial assemblies growing in narrow channels, for example, stress anisotropy and boundary geometry align elongated cells with the expansion flow, and division axes become biased by the mechanically selected director field \cs\cite{hupePredictionControlGeometryinduced,isenseeStressAnisotropyConfined2022}. In other cases, the axis of division can be influenced by local cues such as the extracellular matrix lineage orientation memory, or other physical cellular properties \cs\cite{loyerLastbornDaughterCell2018,gloerichCellDivisionOrientation2017,theryExtracellularMatrixGuides2005a,fineganDivisionOrientationDisentangling2019a}. The division axis can then in turn influence the dynamics of the collective by sustaining a robust spatial organization. Losing control of such division orientation can lead to tissue disorganization\cs\cite{lechlerAsymmetricCellDivisions2005}, disrupt organ development and homeostasis\cs\cite{fischerDefectivePlanarCell2006} and can promote disease progression\cs\cite{peaseMitoticSpindleMisorientation2011} like cancer.

Here, we investigate these questions using a minimal particle-based model, which focuses on two ingredients broadly shared between a range of biological systems: differential growth from nutrient limitation and mechanical interactions through steric repulsion. Nutrients diffuse inward from the boundary and locally control consumption and growth, producing active zones where nutrients are abundant and dormant regions where they are depleted. Guided by an analytical model, we tune the system into a quasi-steady traveling-wave regime and, by varying the nutrient flux, recover the transition from flat, bulk-like fronts to strongly branched, diffusion-limited protrusions. 

Within this minimal setting, we introduce a tunable lineage memory of the division axis, that encodes a minimal rule by which a cell passes its growth axis to its descendants. Rather than modeling a specific biological pathway, this parameter captures the essential effect of persistent polarity, allowing us to interpolate between systems that maintain a stable division axis and systems in which orientation is randomized and reset at each generation. This framework enables us to disentangle how inherited polarity competes with mechanically driven reorientation, showing how growth-generated stresses can reinforce or erase lineage memory and how these interactions depend on the emergent front morphology. In the following, we show how this interplay shapes collective organization in growing assemblies and under what conditions lineage memory meaningfully contributes to large-scale alignment.

\section{Methods}

\begin{figure}[t]
    \centering
    \includegraphics[width=\linewidth]{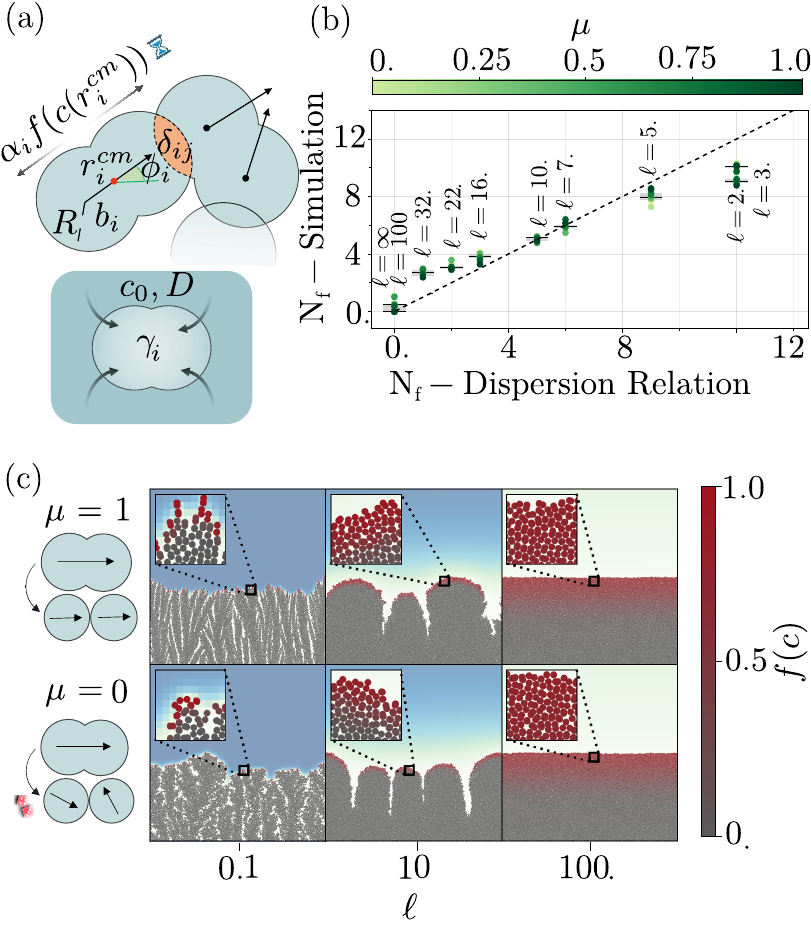}
    \caption{
        \panel{pan:abm-figure}~ Visual description of the agent based model
        \panel{pan:nfing_sim_vs_disp}~Number of protrusions measured in the simulation when they are just formed vs the number of protrusions predicted from the dispersion relation $\Omega(\ell,k)$.
        \panel{pan:snapshots_zoom}~ Snapshots of the system for $\ell=0.1,16.$ and $100.$, colored wrt the growth factor $(f(c))$: Red if $f(c)=1$ (active cells) and Grey if $f(c)=0$ (dormant).
    }
    \label{fig:fig1}
\end{figure}

\subsection{Particle-based framework}

To investigate how growth, mechanical interactions, and division-axis inheritance shape collective organization, we use a particle-based model that extends the mechanically consistent 2D disk-cell framework of Hupe \textit{et al.}\cs\cite{hupeMinimalModelSmoothly} to include nutrient–field interactions, and tunable division axis memory. As shown in \cref{pan:abm-figure} each cell is represented as two overlapping disks of radius $R$ connected by an extensible backbone of length $b_i$. As the internal clock $g_i\in[0,1)$ advances, the backbone relaxes toward its equilibrium length $b_i^{\mathrm{eq}}(g_i) = 2R g_i$ and the age increases according to $\dot{g}_i = \alpha_i f\bigl(c(\mathbf{r}_i^{\mathrm{cm}})\bigr)$
where $\alpha_i$ is the intrinsic growth rate and $f(c)$ is a regulatory function of the external field concentration $c$ evaluated at the center-of-mass position $\mathbf{r}_i^{\mathrm{cm}}$.

Cells consume the external field at a rate 
$\gamma_i f\!\left(c(\mathbf{r}_i^{\mathrm{cm}})\right) A(b_i)$, 
where $A(b_i)$ is the effective projected area of the dumbbell-shaped cell (see \cref{Apndx:ABM}). The external field $c(\mathbf{r},t)$ represents a nutrient concentration defined on a regular grid that spans the simulation domain. This field evolves according to a diffusion–consumption equation so that cells both read out the local value of $c$ to set their growth and, in turn, deplete it through consumption. 

Cell motion is overdamped. The center of mass, backbone length, and orientation respond to repulsive overlap forces, internal elasticity, and torques from neighbouring cells. These forces follow Hertzian contact laws with an effective Young modulus $Y$ that sets the stiffness of steric repulsion. In all simulations we choose $Y$ sufficiently large so that typical overlaps remain small compared to $R$, keeping the colony in a regime that is effectively incompressible at the scale of individual cells (see \cref{Apndx:ABM} and Ref.~\citenum{hupeMinimalModelSmoothly}). Complete force and mobility definitions are provided in \cref{Apndx:ABM}.

When the internal clock reaches its threshold, $g_i = 1$, the parent cell divides into two daughters placed at the positions of its terminal nodes. Mechanical parameters are inherited, while the division axis follows a tunable lineage-memory rule $\varphi_{\mathrm{daughter}}=\varphi_{\mathrm{parent}} + \eta$ where the noise term is sampled as $\eta \sim (1-\mu)\mathcal{U}\left[-\frac{\pi}{2},\frac{\pi}{2}\right]$. Here $\mathcal{U}$ denotes a uniform distribution and the parameter $\mu\in[0,1]$ controls the strength of division-axis memory. For $\mu\to 1$, daughters inherit their parent’s orientation, whereas for $\mu\to 0$ the division orientation is completely random and lineage information is lost upon division. 

All simulations make use of \textit{InPartS}, our open-source framework\cs\cite{inparts} for mechanically consistent particle-based models.

\subsection{Analytical tools}\label{Sec:analytics-main}

To interpret the behavior of the particle-based simulations, we also consider a coarse-grained continuum description that tracks only the colony height and nutrient field, leaving cell orientation aside. At this level we treat the colony as an incompressible, friction-dominated medium at close-packed density $\rho_0$, so that in the Darcy limit the velocity field satisfies $\mathbf{u} = -\nabla p / \xi$. Mass conservation with growth then gives

\begin{equation}\label{eq:mass-cons}
\nabla \cdot \mathbf{u} = \alpha_0 f(c),
\end{equation}
where proliferation is regulated by a Monod factor $f(c) = c / (c + c_h)$ that captures growth suppression at low nutrient.

The nutrient concentration $c(y,t)$ is supplied from above at a fixed value $c_b$ and is consumed only inside the colony. In one dimension normal to the front, it obeys
\begin{equation}\label{eq:reac-dif}
\partial_t c = D\,\partial_y^2 c
- \gamma \rho_0 f(c)\, \Theta\bigl(h(t) - y\bigr),
\end{equation}
where $D$ is the diffusion coefficient, $\gamma$ is the consumption rate per unit biomass, and $h(t)$ is the front position. Far above the colony diffusion equilibrates the nutrient, so the gradient vanishes and $c \to c_b$.

The advancing colony front acts as a moving sink, producing a depletion zone just above it. When diffusion cannot equilibrate the domain, regions far ahead remain unaffected and the nutrient gradient vanishes. To analyze this quasi-steady regime, we follow Ref. \citenum{wangShapeGrowingFront2017} and seek traveling-wave solutions $c(y,t) = \hat c(z)$ in the comoving coordinate $z = y - v t$ with constant front speed $v$. This reduces the nutrient dynamics to a one-dimensional reaction–diffusion problem in $z$ that can be solved with appropriate continuity and flux conditions at the front (see \cref{Apndx:Analytics}).

Three length scales emerge from this analysis: the diffusion–consumption length $\ell_D = \sqrt{D c_b / (\gamma \rho_0)}$, which measures how far growth can significantly deplete the nutrient; the depletion-zone thickness $\ell_d = D / v$, which sets the size of the region ahead of the front where the nutrient profile is markedly distorted by the moving sink; and the active-layer width $\ell_a = v c_b / (\gamma \rho_0) = v / \alpha_0$, which quantifies the thickness of the region inside the colony where growth remains appreciable.

They satisfy the identity $\ell_D^2 = \ell_d \ell_a$ (derived in the Appendix). In the limit $c_h / c_b \equiv \varepsilon \ll 1$ the transition between the active and starved regions becomes sharp. In this limit the growth response becomes a sharp switch, with $f = 1$ in a thin active layer just behind the front and $f = c / c_h$ deeper in the depleted interior, which makes \cref{eq:reac-dif} analytically tractable. The resulting solution depends on the dimensionless parameter $\lambda = \ell_D / \ell_d = \ell_a / \ell_D$, which is not fixed by the matching conditions but can be shown to satisfy $\varepsilon^{1/2} < \lambda < (2\varepsilon)^{-1/2}$ (see \cref{Apndx:Analytics}). For the parameters used here ($\varepsilon = 0.01$), this implies $\lambda = O(1)$, so that the active-layer width and diffusion–consumption length are comparable, $\ell_a \sim \ell_D$. In the following we therefore refer to this common scale simply as $\ell$, and use $\ell$ to denote both the diffusion–consumption length and, by extension, the active-layer thickness.

To determine when a constant-velocity traveling wave can be sustained, we follow the approach of Farrell \textit{et al.}\cs\cite{farrellMechanicallyDrivenGrowth2013}. By integrating the traveling-wave form of \cref{eq:reac-dif} over the whole domain and using the mass-conservation relation \cref{eq:mass-cons}, we obtain the constraint
\begin{equation}\label{beta-constraint}
    \beta \equiv \frac{c_b Y}{\rho_0} = 1,
\end{equation}
where $Y = \alpha_0 / \gamma$ is the yield, that is, the biomass produced per unit nutrient consumed. The dimensionless parameter $\beta$ therefore measures the potential biomass that can be generated from the boundary nutrient, normalized by the maximal packing density. Keeping $\beta \approx 1$ selects a quasi-steady traveling-wave regime in which the far field acts as a time-independent reservoir and only local fluctuations near the front can feed back on growth and flow.

Within this reduced description, pattern formation such as fingering is thus governed by the local physics of the nutrient-limited front. In the next step we perturb the flat traveling wave and perform a linear stability analysis to obtain a dispersion relation $\Omega(k,\ell)$ for small front undulations. This allows us to determine which lateral modes grow in time and to estimate the characteristic number and spacing of protrusions as functions of the diffusion length $\ell$ and the half-saturation concentration $c_h$.

\section{Results}

\begin{figure}[t]
    \centering
    \includegraphics[width=\linewidth]{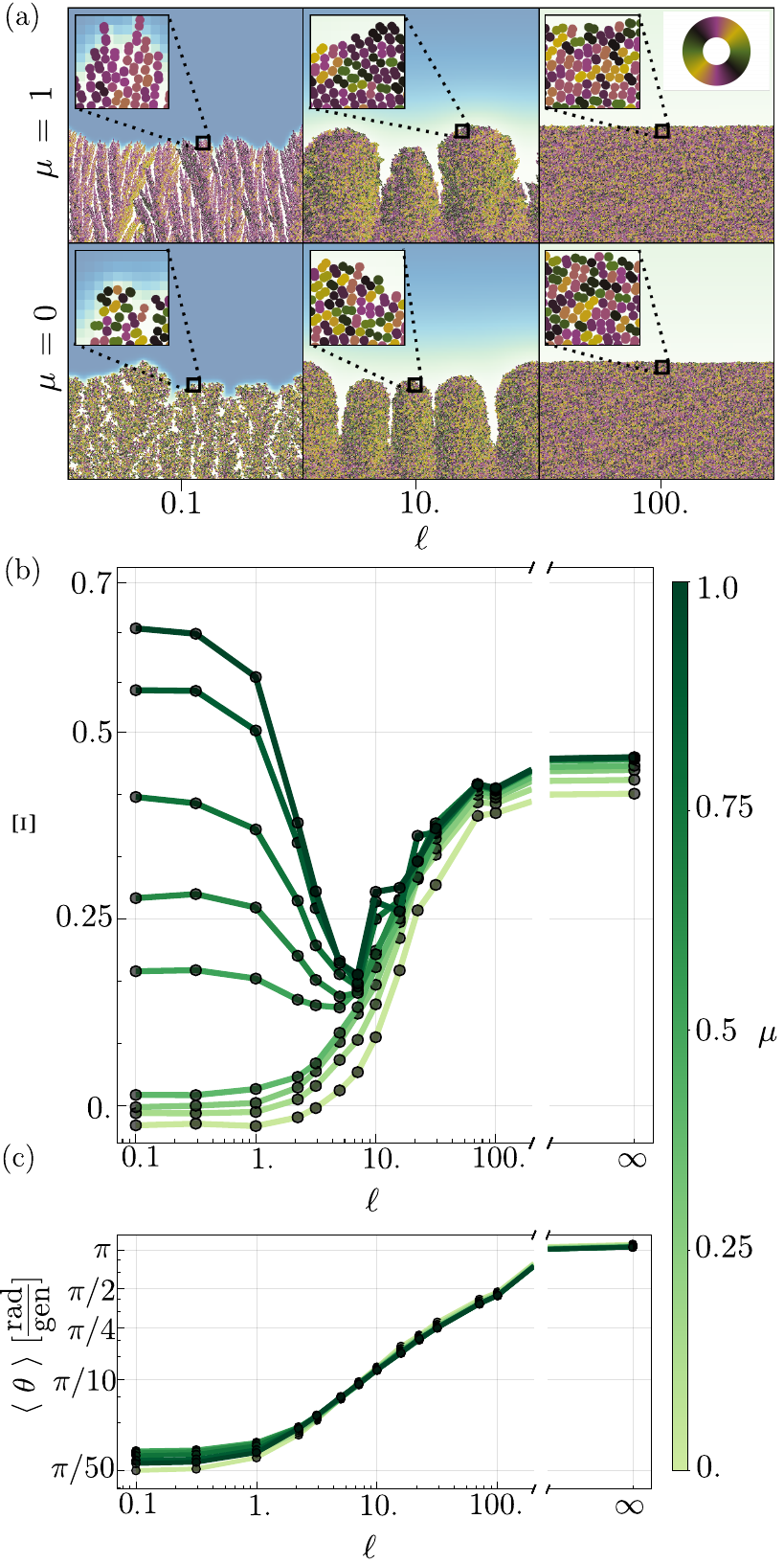}
    \caption{
        \panel{pan:snapshots_zoom_orient}~Snapshots for $\ell = 0.1, 16, 100$ with cells colored by their orientation according to the color wheel.
        \panel{pan:xi_vs_ell}~Global orientational order $\Xi$ as a function of $\ell$.
        \panel{pan:rotvel_ell}~Average rotational velocity of cells with $f(c) > 0.1$ for different values of $\ell$.
    }
    \label{fig:fig2}
\end{figure}

\subsection{Simulation setup: colonies under nutrient limitation}
To probe how nutrient transport and lineage memory shape front morphology in a controlled setting, we simulate growth in a 2D domain. We take the cell diameter as the basic length unit and choose a domain of height $L_y$ and width $L_x = 200$ with periodic boundary conditions along $x$. The nutrient concentration field is defined on a regular grid covering this domain and is initially uniform at the boundary value $c_b = 1$, which sets the concentration scale. The initial colony consists of a single row of $100$ cells placed at the bottom boundary with random orientations and identical mechanical parameters, including a large effective Young modulus $Y = 10^6$ to keep overlaps small and the packing effectively incompressible.

Growth and consumption follow the Monod regulation $f(c) = c/(c + c_h)$ with $c_h = 0.01$. At each division, the daughter’s intrinsic growth rate $\alpha_i$ is redrawn to avoid collective phase-locking of the cell cycle, sampling $\alpha_i \sim \mathcal{U}\bigl[\alpha_0 - \alpha_0/4,\; \alpha_0 + \alpha_0/4\bigr]$ with $\alpha_0 = 1$, so that one time unit corresponds to the mean cell-cycle duration. The consumption rate is then chosen as $\gamma_i = c_b / \alpha_i$ which maintains the constant-velocity front condition of \cref{beta-constraint}. The memory parameter $\mu \in [0,1]$ tunes the strength of division-axis inheritance.

We explore a broad range of nutrient-limited morphologies by varying the diffusion coefficient $D \in \{0.01, 0.1, 1, 5, 10, 25, 50, 100, 250, 500, 1000, 3000, 5000, 10^4\}$ and $D\to\infty$ along with a range of lineage-memory parameter $\mu \in \{0, 0.3, 0.5, 0.6, 0.8, 0.85, 0.9, 0.95, 1\}$. The case $D \to \infty$ corresponds to homogeneous growth with no nutrient field or consumption. The snapshot time step is $\Delta t = 1/(\alpha_016) = 1/16$, and each $(D,\mu)$ pair is simulated with $20$ independent realizations.

To keep the simulations efficient while faithfully resolving the active front, we combine the particle-based model with the continuum analysis of \cref{Sec:analytics-main}. First, the nutrient profile from the traveling-wave solution is used to determine a height $H$ above the colony where the concentration differs from the boundary value by at most $\lvert c(H) - c_b \rvert = \delta c = 0.01$ (\cref{Apndx:Analytics}). Above this height we fix $c = c_b$, avoiding the need to simulate a large, essentially uniform reservoir. Second, we exploit the fact that cells with $f(c) < 10^{-3}$ are effectively dormant: they neither grow nor move appreciably and sit in a region where $c \to 0$. At regular intervals we identify all cells with $f(c) > 10^{-3}$, determine the lowest vertical position among them, and take this as the back edge of the active layer. Cells in a strip of fixed width (5 cell diameters) just behind this edge are frozen and kept as a mechanically attached scaffold, while cells deeper in the interior are removed and exported for later analysis. In this way the simulation focuses computational effort on the actively growing layer while still retaining information about the colony interior.

As $D$ increases, the diffusion length $\ell$ and the active-layer width grow, requiring a coarser spatial resolution for the nutrient field. To balance accuracy and cost, we adjust the diffusion grid spacing $\delta x$ as a function of $D$: we use $\delta x = 1$ for $D \le 500$, $\delta x = 2$ for $500 < D \le 1000$, and $\delta x = 4$ for $1000 < D \le 10^4$. For $D \ge 5\times 10^4$ we are in the homogeneous limit and do not solve the diffusion equation.

By tuning $D$ in this way we access a continuum of front morphologies, from strongly branched, diffusion-limited patterns to nearly flat, bulk-like fronts as the active layer widens (larger $\ell$). This is illustrated in \cref{pan:snapshots_zoom}, where the top row shows snapshots for $\mu = 1$ and the bottom row for $\mu = 0$, with cells colored by their growth factor (red for $f \to 1$ and gray for $f \to 0$). Despite varying the division-axis memory, the emergence of the distinct morphological patterns is preserved, meaning that is controlled primarily by nutrient-limited growth and steric mechanics rather than by lineage inheritance. Instead, lineage memory refines the structure of individual fingers: for strong memory, thin protrusions remain narrow and pointed, whereas for $\mu = 0$ the same fronts develop more ragged, noisy tips.

To quantify this, we measure the number of protrusions that appear after the transient. For each simulation we identify the outermost cells along the front, construct their height profile as a function of lateral position, and compute its Fourier spectrum. The dominant peak yields the characteristic wavenumber of the perturbed front and thus the number of fingers. We then compare this to the theoretical prediction obtained from the dispersion relation $\Omega(k,\ell)$ derived from the continuum model (\cref{Apndx:Analytics}). As shown in \cref{pan:nfing_sim_vs_disp}, the measured protrusion number closely matches the theoretical estimate across the range of $\ell$, confirming that the instability and its selected scale are set by the nutrient–mechanical front physics and are largely insensitive to the division-axis memory parameter $\mu$.

\subsection{Large-scale orientational patterns}
We have seen that the emergence of branched versus flat fronts is robust to changes in division-axis memory. We now turn to how lineage inheritance and nutrient-limited mechanics jointly organize the packing of the colony, and in particular its large-scale orientational order. To visualize this, we color each cell by its nematic orientation (\cref{pan:snapshots_zoom_orient}). For a quantitative measure we use a global nematic order parameter $\Xi = \bigl\langle \cos\bigl[2(\varphi_i - \pi/2)\bigr] \bigr\rangle$
averaged over actively growing cells. This quantity approaches $+1$ when cells are predominantly vertical, $-1$ when they are horizontal, and $0$ when orientations are isotropically distributed.

For each pair $(\ell,\mu)$ we evolve the system until $\Xi$ reaches a statistically stationary regime and then time-average over that window. The resulting dependence of $\Xi$ on $\ell$ is shown in \cref{pan:xi_vs_ell} for several values of the lineage memory $\mu$. To disentangle the contribution of inheritance from that of mechanically driven reorientation, we also compute the mean angular velocity of actively growing cells, obtained by finite differences of $\varphi_i(t)$ between successive snapshots and averaging over cells with $f(c) > 0.1$ (\cref{pan:rotvel_ell}).

In the strongly diffusion-limited regime (small $\ell$, thin active layer) colonies develop narrow fingers. For $\mu = 1$ these fingers are sharply organized: cells are almost perfectly vertically aligned and $\Xi$ is close to $1$, while for $\mu = 0$ the same morphologies display nearly isotropic orientations with $\Xi \approx 0$. At the same time, angular velocities remain small for all $\mu$, indicating that cells barely rotate during their lifetime. In this regime the single-cell growth axis is therefore largely decoupled from the propagation direction of the front: the front advances along $y$, but cells retain the division axis they inherited, and this axis sets the direction in which they primarily exert forces.  Together, these observations show that in this regime orientational order is dominated by inherited division axes, with mechanics playing only a minor role in reorienting cells once they are born.

As $\ell$ increases to intermediate values, the active layer thickens and fingers broaden. More cells grow simultaneously within each protrusion, stresses become larger and more heterogeneous, and mechanical interactions begin to rotate cells away from their inherited orientation. The fronts develop chevron-like orientational patterns. In this regime $\Xi$ decreases for $\mu = 1$ but increases for $\mu = 0$, reflecting the fact that mechanical reorientation now competes with lineage memory: it weakens the vertically ordered state that inheritance tries to maintain, yet it also imposes structure on an otherwise disordered orientation field when $\mu$ is small.

For very large $\ell$ (thick active layers and nearly flat fronts), mechanical effects dominate. A large fraction of cells are active and strongly coupled, and the coherent flow generated by growth reorients them toward a vertically ordered configuration. As $\ell$ increases, the mean angular velocity grows and then saturates at a value of order $\pi$ per cell cycle, indicating that cells undergo a finite reorientation before settling into a mechanically selected alignment that is no longer strongly reshaped by subsequent dynamics. In this regime the nematic order parameter rises again for all $\mu$, showing that alignment is maintained primarily by force and flow-driven reorientation rather than by inheritance alone. Here single-cell growth axes become effectively locked to the macroscopic $y$-directed advance of the front, in contrast to the fingering regime where growth and front propagation directions remain decoupled.Varying $\ell$ thus drives a crossover from a memory-dominated regime with vertically aligned fingers, through an intermediate regime where inheritance and mechanics interfere and reduce global order, to a force-dominated regime where mechanical interactions generate strong large-scale alignment even when lineage memory is weak.

\begin{figure}[t]
\centering
\includegraphics[width=\linewidth]{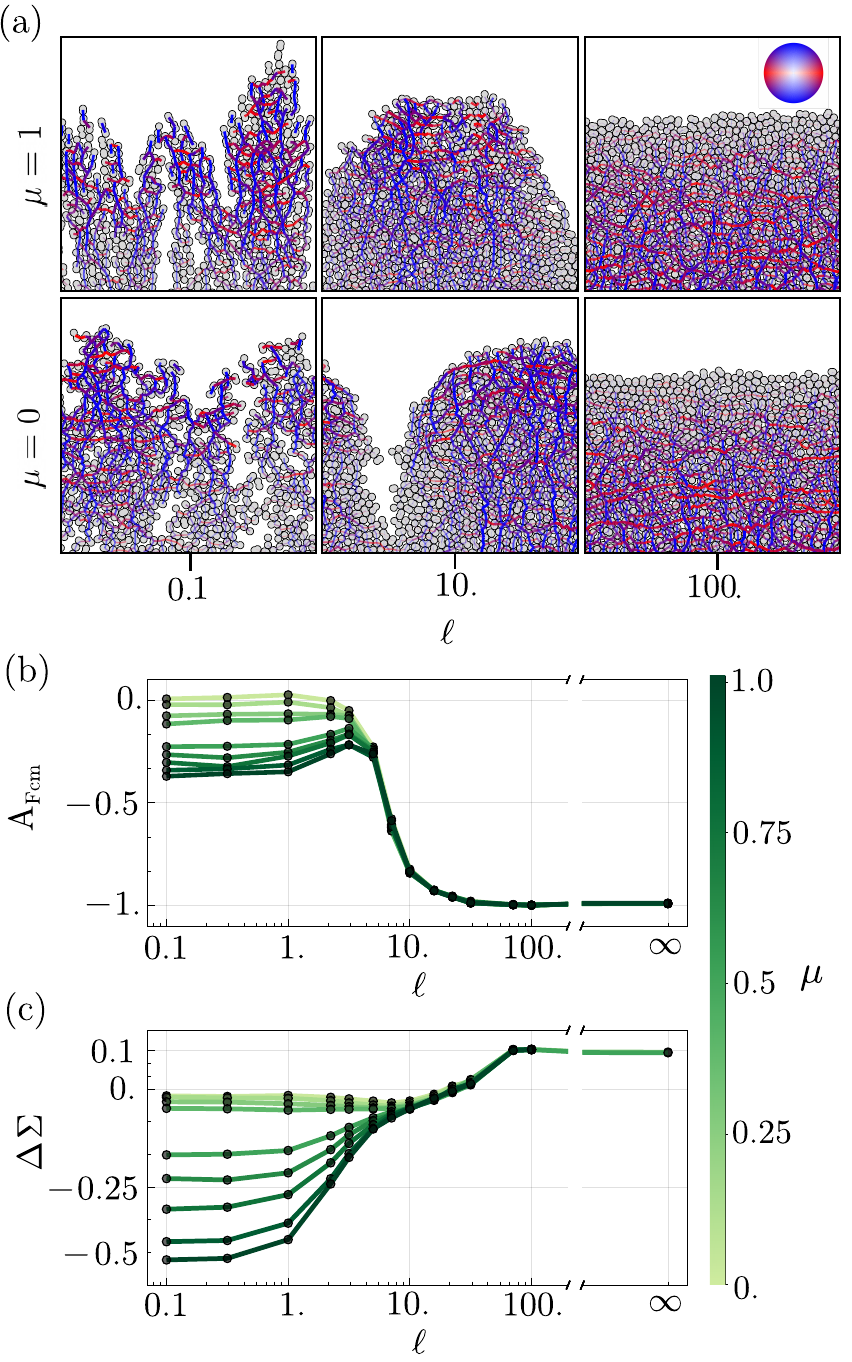}
    \caption{
        \panel{pan:snapshots_dipoles}~Contact-force maps for $\ell = 0.1, 16,$ and $100$, showing node–node interaction forces as oriented line segments colored by force direction and scaled by magnitude.
        \panel{pan:asymmetry_vel}~Center-of-mass force asymmetry $A_{F_{\mathrm{cm}}}$ as a function of $\ell$ for different values of the division-axis memory $\mu$.
        \panel{pan:stress_anisotropy}~Stress anisotropy $\Delta\Sigma$ versus $\ell$, computed from the coarse-grained Cauchy stress tensor for the same set of $\mu$.}
    \label{fig:fig3}
\end{figure}

\subsection{Mechanical dynamics}
Having identified a crossover from memory-dominated to force-dominated alignment, we next ask how this transition is reflected in the collective mechanics of the colony. To do so, we exploit the fact that the particle-based model provides the full set of pairwise interaction forces between nodes as well as the net force acting on each cell’s center of mass. From these data we construct force-map visualizations (\cref{pan:snapshots_dipoles}) in which each contact is drawn as a short line segment oriented along the force direction; the color hue encodes the angle (horizontal forces appear red, vertical forces blue, with smooth transitions in between), while opacity and line width increase with force magnitude. To make the contrast robust to outliers, we plot the logarithm of the force magnitude, clipped to the 5th–95th percentiles and rescaled to the interval $[0,1]$ so that strong contacts stand out clearly while very weak ones remain faint.

Beyond these visualizations, we quantify two complementary measures of mechanical anisotropy. The first is a center-of-mass force asymmetry, $A_{F_{\mathrm{cm}}} = \left\langle (\lvert F_x \rvert - \lvert F_y \rvert) / (\lvert F_x \rvert + \lvert F_y \rvert) \right\rangle$, where $F_x$ and $F_y$ are the horizontal and vertical components of the net force on each cell and the average is taken over actively growing cells. This quantity lies in $[-1,1]$, with negative values indicating predominantly vertical forcing, positive values predominantly horizontal forcing, and values near zero indicating no preferred direction.

The second measure is a stress anisotropy, $\Delta \Sigma = (\lvert \sigma_{xx} \rvert - \lvert \sigma_{yy} \rvert)/(\lvert \sigma_{xx} \rvert + \lvert \sigma_{yy} \rvert)$, where $\sigma_{xx}$ and $\sigma_{yy}$ are the normal components of a coarse-grained Cauchy stress tensor constructed from contact interaction forces $\boldsymbol{\delta} \otimes \mathbf{f}$, with $\boldsymbol{\delta}$ the branch vector between interacting nodes and $\mathbf{f}$ the corresponding interaction force (see \cref{Apndx:ABM}). Like $A_{F_{\mathrm{cm}}}$, $\Delta \Sigma \in [-1,1]$, but it characterizes how normal loads are distributed (vertical versus horizontal) rather than how cells are instantaneously driven to move.

The behavior of these two quantities across front morphologies is summarized in \cref{pan:asymmetry_vel} and \ref{onlyletter:pan:stress_anisotropy}. In the strongly fingering regime (small $\ell$), colonies with strong lineage memory ($\mu = 1$) exhibit a clear vertical bias: $A_{F_{\mathrm{cm}}} < 0$ and $\Delta \Sigma < 0$, showing that cells are both driven and stressed predominantly along the growth direction. For $\mu = 0$, by contrast, both measures are close to zero, consistent with the nearly isotropic orientations seen in \cref{pan:snapshots_zoom_orient}. In this regime, mechanical interactions do not on their own generate a directional bias; instead, they primarily transmit the inherited division axis when that memory is present.

As $\ell$ increases and the active layer thickens, both measures evolve in a nontrivial way. The center-of-mass force asymmetry grows towards $A_{F_{\mathrm{cm}}} \approx -1$, indicating that cell motion becomes almost purely vertical, while the stress anisotropy $\Delta \Sigma$ turns positive. This combination signals a mechanically confined state: growth-driven vertical flows push cells upward, but the resulting crowding generates strong lateral normal stresses so that the colony increasingly bears its mechanical load in the horizontal direction. Although velocities and center-of-mass forces show a strong $y$-bias in this regime, the underlying dipolar contact forces and stresses are skewed toward $x$, revealing how vertical growth and front advance are sustained by lateral load-bearing. In this force-dominated regime the dominant role of mechanics is no longer to simply transmit inherited polarity, but to reshape it, imposing a collective stress pattern that ultimately drives reorientation to release stress.

\section{Competition dynamics}
\begin{figure}[t]
    \centering
    \includegraphics[width=\linewidth]{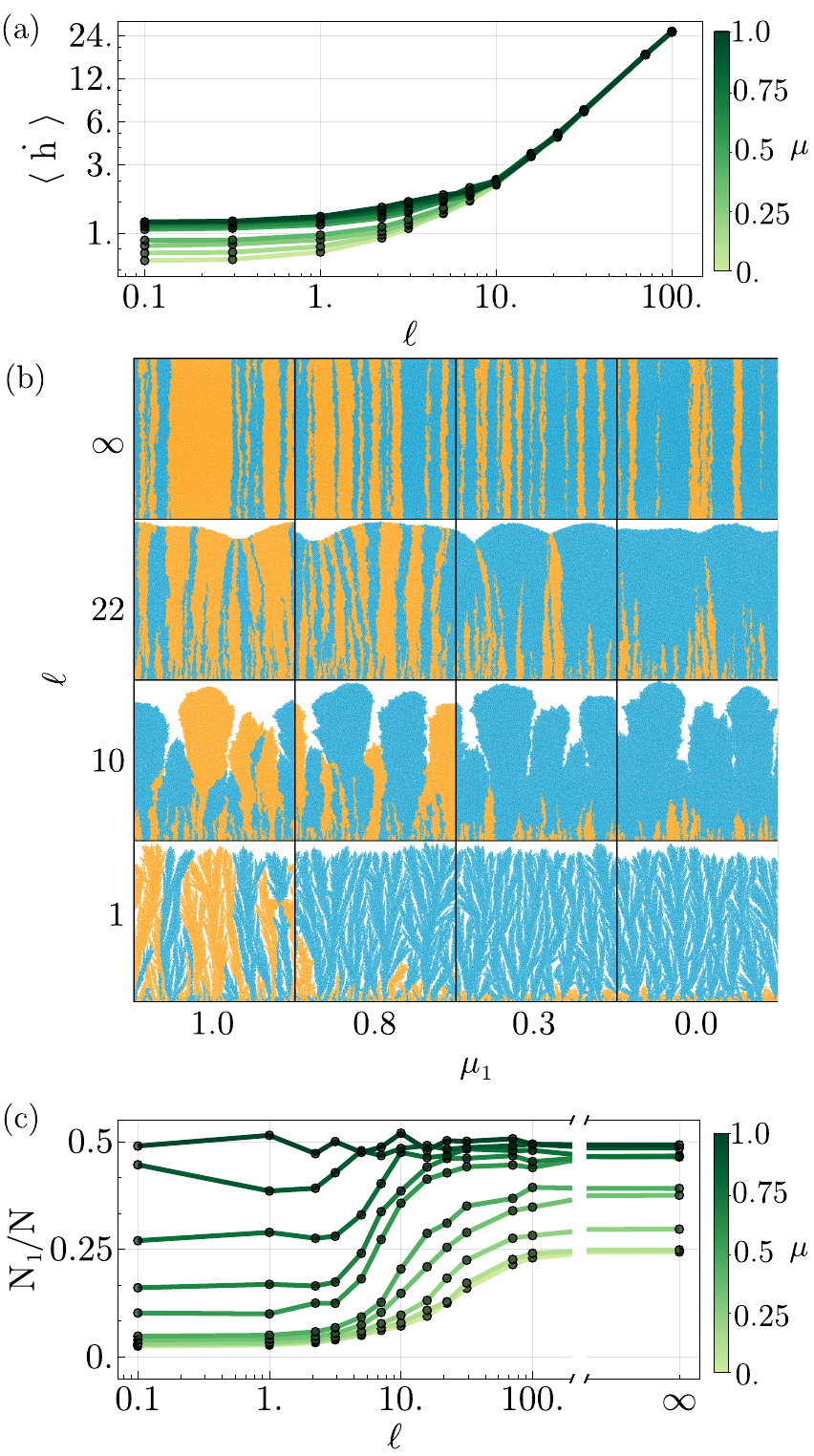}
    \caption{
        \panel{pan:Velocity-front}~Front propagation speed $v$ as a function of the diffusion length $\ell$ for different values of the division-axis memory $\mu$.
        \panel{pan:phase-diagram-comp}~Representative final configurations from two-species competition for selected $(\ell,\mu_1)$, with species~1 (orange, variable $\mu_1$) competing against species~2 (blue, $\mu_2 = 1$).
        \panel{pan:competition-diagram}~Fraction of species~1, $N_1/N$, at the final time, summarizing competition outcomes as a function of $\ell$ and $\mu_1$.
    }
    \label{fig:fig_comp}
\end{figure}
We have seen that in the fingering regime with strong lineage memory ($\mu = 1$) cells are almost perfectly vertically aligned, even though the stress anisotropy does not impose a corresponding vertical bias. To clarify the origin of this alignment, we first examine the front kinematics. In \cref{pan:Velocity-front} we measure the front speed by identifying cells at the leading edge, computing their mean vertical position, and extracting the corresponding propagation velocity. As the diffusion length $\ell$ increases, the front speed grows and collapses onto a single curve for different values of $\mu$, except in the strongly diffusion-limited regime. There, fronts with $\mu = 1$ propagate systematically faster than those with $\mu = 0$, indicating a small but robust fitness advantage for lineages whose growth axis is aligned with the direction of advance and nutrient influx.

We probe this advantage more directly using competition simulations between two otherwise identical species that differ only in their division-axis memory. Species 2 (blue) has $\mu_2 = 1$, whereas species 1 (orange) has a tunable memory $\mu_1 < 1$. The initial condition is a $50$–$50$ mixture of the two types arranged randomly within a narrow strip at the base of the domain. The colony then grows until the front reaches a height of order $300$ cell diameters. For each pair $(\ell,\mu_1)$ we perform $40$ independent realizations. Representative final configurations are shown in \cref{pan:phase-diagram-comp} as a function of $\ell$ and $\mu_1$, and the corresponding statistics are summarized in \cref{pan:competition-diagram} via the mean orange fraction $N_1/N$ at the final time.

In the fingering regime (small $\ell$), the outcome depends strongly on $\mu_1$. When the orange species has $\mu_1 \to 0$, the blue $\mu_2 = 1$ cells almost always dominate the advancing edge; as $\mu_1$ increases toward $1$, the two species tend toward coexistence with comparable abundances. These results demonstrate that, in diffusion-limited fronts, orientational memory enhances access to fresh nutrient and thereby increases effective growth rate. Vertical alignment in fingers with $\mu = 1$ is thus not imposed directly by stress, but is selected by growth: lineages whose division axes point toward the nutrient source advance more rapidly and, through inheritance, progressively take over the front.

In contrast, for large $\ell$ (flat fronts with thick active layers) the two species generically coexist across the full range of $\mu_1$, with only a mild bias in favor of the high-memory species at very low $\mu_1$. In this regime the nutrient field is more homogeneous and growth is constrained by mechanical confinement in the dense bulk, so nutrient access no longer depends sensitively on single-cell orientation. Instead, it is controlled by the collective flows and stress patterns generated by the expanding front, consistent with the force-dominated alignment regime identified above.

\section{Discussion and outlook}

Our results disentangle the rules that control front morphology in nutrient-limited growth from those that self-organize cell orientations and flows, and further isolate how oriented division through lineage memory feeds back onto these collective dynamics. Using a continuum traveling-wave description coupled to a particle-based model, we found that the emergence of flat versus fingered interfaces is controlled primarily by nutrient-limited growth. The oriented division rule does not shift this morphological transition, but it reorganizes the active region so that fingers may or may not develop coherent nematic order, with direct consequences for propagation speed and lineage competition.

In the regime where mechanical interactions only weakly reorient cells over a cell cycle, the single-cell growth axis in our simulations is largely decoupled from the propagation direction of the front. The colony as a whole advances along the nutrient gradient, yet individual lineages can maintain their own growth orientation locally. This decoupling is analogous to systems that exhibit branching morphogenesis, where models have described the terminal tips as branching random walkers, where local branching angles are effectively stochastic to enhance spatial exploration or boundary length, yet the composite branched network remains biased toward an overall direction of growth\cs\cite{Hannezo2017-fl,Ucar2021-nv}. 

At the same time, the main role of lineage-based orientation memory in our model is to allow adaptation especially when other cues (like mechanics) cannot impose their own order. In this diffusion-limited regime, inherited division axes allow lineages to collectively maintain a growth direction that is beneficial, in our case, pointing toward the nutrient source so that cell elongation and division do productive work pushing the front, thereby increasing access to fresh resource and front speed. In this sense, oriented division drives the mechanics. This scenario similar to systems of filamentous \textit{Saccharomyces cerevisiae}, where nutrient limitation induces pseudohyphal and invasive growth as chains of cells that sprout from the colony edge by repeated polarized budding along one axis. In those colonies, the fact that mothers and daughters keep dividing along a shared axis creates coherent protrusive strands that dig into fresh substrate and are viewed as a foraging advantage when nutrients are scarce \cite{FilamentousgrowthSandra2004}. In more complex systems such as Drosophila larval neuroblasts that are ensheathed by cortex glia\cs\cite{loyerLastbornDaughterCell2018}, division-axis memory plays a different but related role: successive asymmetric divisions are repeatedly oriented relative to the neuroblast–glia interface so that each division produces a self-renewing neuroblast that remains in contact with the glial niche and a differentiating daughter that is displaced away from the interface. In this way, lineage-based memory of the division axis preserves a beneficial spatial configuration across many generations, where neuroblasts anchored at the niche, daughters positioned outward, especially in such scenarios where growth is highly localized and cell rearrangements alone are unlikely to robustly organize the tissue.

At the opposite end of parameter space, when the active layer becomes thick and confinement strong, our simulations enter a regime in which mechanical interactions dominate the orientational dynamics. Cells then undergo substantial rotation during a cell cycle and eventually align with the collective direction of motion, independently of their lineage history. The contact-force and stress fields reveal vertical motion supported by horizontal load-bearing, consistent with a mechanically confined expansion, similar to previously studied systems of confined bacterial growth in channels, where rod-shaped bacteria self-organize into ordered patterns that are strongly constrained by geometry and stress anisotropy\cs\cite{isenseeStressAnisotropyConfined2022}. Hupe \textit{et al.} demonstrated that, under isotropic growth, the shear rate of the expansion flow and the shape of the confining domain already predict the orientation patterns in dense colonies\cs\cite{hupePredictionControlGeometryinduced}. Our thick-layer regime can be viewed as a related situation without hard walls: the effective confinement arises from growth against a dense bulk, and the orientational field is selected by the expansion flow and stress distribution. 

Taken together, these results suggest a way to interpret orientational patterns in growing assemblies. If, in an unconfined expanding system, local division axes remain aligned with the direction of advance of protruding regions over several generations, this may indicate that mechanical interactions alone are not sufficient to randomize orientations on that timescale and that some form of persistent orientational bias or memory helps to maintain a coherent structure. Conversely, when the dominant cell and division orientations closely track confining boundaries, large-scale expansion flows, or principal stress directions, as in growth within narrow channels or strongly constrained colonies, it is likely that mechanical effects are the main drivers of organization and might overwrite or work together with any other mechanisms of oriented division. Therefore, our study provides a language for relating measured orientation fields to underlying growth and interaction regimes.

More generally, our approach isolates growth and division-axis inheritance as explicit control parameters in a setting where, in real biological systems, they are often entangled with other forms of activity. In the present work, we implemented only one concrete version of lineage memory, so how robustly similar regimes emerge under alternative microscopic mechanisms of orientational persistence remains to be explored. Nonetheless, we expect the distinction between regimes in which oriented growth drives mechanics and regimes in which collective mechanical effects reinforce or reshape oriented growth and alignment to be a useful lens for interpreting experiments on fingering, branching, and invasive growth in colonies, tissues, and organoids, many of which likely reside somewhere along this spectrum but for which the respective roles of growth rules and mechanics are not clearly disentangled.\\

\acknowledgments
We thank Jonas Isensee and Lukas Hupe for valuable discussions, input and technical help with the simulations. We acknowledge support from the Max Planck Society as well as the Max Planck School Matter to Life, which is jointly funded by the Federal Ministry of Education and Research (BMBF) of Germany and the Max Planck Society.

\interlinepenalty=10000

\bibliographystyle{apsrev4-2}

\appendix

\begin{figure}[h]
    \centering
    \includegraphics[width=\linewidth]{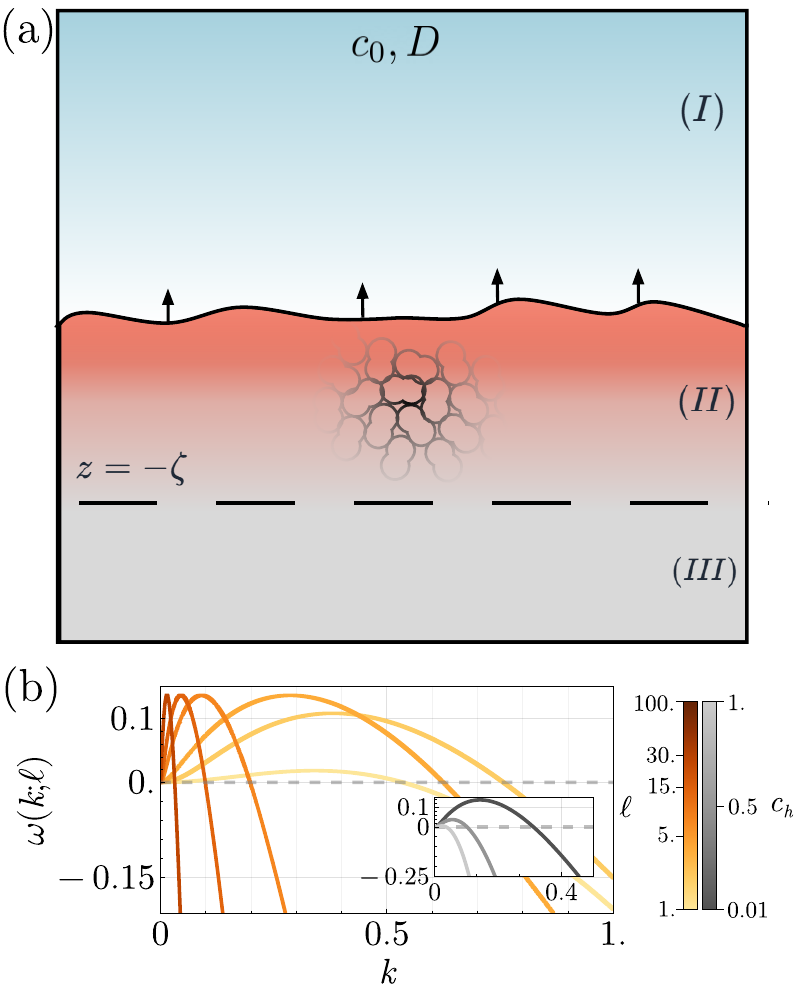}
    \caption{
        \panel{pan:system}~ System
        \panel{pan:dispersion_diagram}~Dispersion diagram $\Omega(\ell,k)$ 
    }
    \label{fig:fig_sup}
\end{figure}

\section{Mechanical details of the particle based model}\label{Apndx:ABM}

The effective area used in the consumption term is
\begin{equation}
A(b_i) =
2\pi R^2
- 2 R^2 \arccos\left(\frac{b_i}{2R}\right)
+ \frac{b_i}{2} \sqrt{4R^2 - b_i^2}.
\end{equation}

Cell motion is overdamped. The center-of-mass $\mathbf{r}_i^{\mathrm{cm}}$, backbone length $b_i$, and orientation $\varphi_i$ evolve according to
\begin{align}
    &\frac{d}{dt}\mathbf{r}_i^{\mathrm{cm}} =
        \chi^\|(b_i) F_i^\| \hat{\mathbf{e}}_i
        + \chi^\perp(b_i) F_i^\perp (\hat{\mathbf{z}}\times \hat{\mathbf{e}}_i),\\
            &\frac{d}{dt} b_i = \chi^{\mathrm{int}}(b_i) F_i^{\mathrm{int}}, 
    \qquad
    \frac{d}{dt}\varphi_i = \chi^{\mathrm{rot}}(b_i) T_i ,
\end{align}
where $\hat{\mathbf{e}}_i$ is the backbone unit vector, $F_i^\|$ and $F_i^\perp$ are the longitudinal and transverse interaction forces, $F_i^{\mathrm{int}}$ is the internal spring force, and $T_i$ is the torque.

Pairwise interactions between nodes of different cells are purely steric and repulsive whenever they overlap. For two nodes $\alpha$ and $\beta$ belonging to cells $i$ and $j$ we define the displacement and overlap as
\begin{align}
    &\mathbf{d}_{ij}^{\alpha\beta} = \mathbf{r}_j^\beta - \mathbf{r}_i^\alpha, \qquad d_{ij}^{\alpha\beta} = \lvert \mathbf{d}_{ij}^{\alpha\beta} \rvert \\
    &\delta_{ij}^{\alpha\beta} = \max\bigl(0, 2R - d_{ij}^{\alpha\beta}\bigr),
\end{align}

For later use in the stress calculation, each interacting node pair $(i,\alpha)$ and $(j,\beta)$ is associated with a branch vector and a contact force,
\begin{equation}
    \boldsymbol{\delta}_{ij}^{\alpha\beta}
    = \mathbf{d}_{ij}^{\alpha\beta}, 
    \qquad
    \mathbf{f}_{ij}^{\alpha\beta}
    = \mathbf{F}_{ij}^{\alpha\beta},
\end{equation}
so that the microscopic contribution of that contact to the Cauchy stress is proportional to the dyadic product
$\boldsymbol{\delta}_{ij}^{\alpha\beta} \otimes \mathbf{f}_{ij}^{\alpha\beta}$.
Summing these dipoles over all contacts and dividing by an appropriate coarse-graining area yields the coarse-grained stress tensor $\boldsymbol{\sigma}$ used in the main text.

When the overlap between 2 cells $\delta_{ij}^{\alpha\beta} > 0$ the contact force follows a Hertzian law and grows with the overlap with exponent $3/2$,
\begin{equation}
    \mathbf{F}_{ij}^{\alpha\beta}
    = m_{ij}\, \frac{Y}{2}\, \sqrt{\frac{R}{2}}\,
      \bigl(\delta_{ij}^{\alpha\beta}\bigr)^{3/2}\,
      \hat{\mathbf{d}}_{ij}^{\alpha\beta}
\end{equation}
and $\mathbf{F}_{ij}^{\alpha\beta} = \mathbf{0}$ otherwise. Here $Y$ is an effective Young's modulus that sets the stiffness of steric repulsion. Large values of $Y$ strongly penalize overlap so that typical overlaps remain small compared to $R$ and the dynamics stays close to an incompressible contact regime. The dimensionless factor $m_{ij}$ is a softness prefactor that is reduced immediately after division in order to compensate for the instantaneous doubling of nodes.

The two nodes of the same cell are coupled by an internal Hertzian spring that tends to relax the backbone length $b_i$ toward its equilibrium value $b_i^{\mathrm{eq}}(g_i,R)$. We write the deviation from this rest length as
\begin{equation}
    \Delta b_i = b_i^{\mathrm{eq}}(g_i,R) - b_i,
\end{equation}
and introduce an internal hardness parameter $H^{\mathrm{int}} > 0$ that controls how rigidly the backbone resists compression or extension. The internal force on node $\alpha$ due to node $\beta$ of the same cell is
\begin{equation}
    \mathbf{F}_{ii}^{\alpha\beta}
    = H^{\mathrm{int}}\, \frac{Y}{2}\, \sqrt{\frac{R}{2}}\,
      \operatorname{sgn}(\Delta b_i)\,
      \lvert \Delta b_i \rvert^{3/2}\,
      \hat{\mathbf{d}}_{ii}^{\alpha\beta}
    \qquad
    (\alpha \ne \beta).
\end{equation}
With $H^{\mathrm{int}} = 1$ the internal spring and intercellular Hertzian forces are chosen so that forces remain continuous as a compressed mother cell divides and its two nodes become the centers of two daughter cells. A detailed derivation and discussion of these force laws can be found in Ref.~\citenum{hupeMinimalModelSmoothly}.

\section{Non-dimensionalization and base state}
\label{Apndx:Analytics}

To analyze the traveling--wave regime, we assume a constant front speed
$v$ and shift to the comoving coordinate $z = y - vt$. Writing
$c(y,t) = \hat c(z)$ yields
\begin{equation}
 -v\, \hat c'(z)
 = D\, \hat c''(z) - \gamma \rho_0 f\bigl(\hat c(z)\bigr)\,\Theta(-z),
 \label{eq:tw-dimensional}
\end{equation}
with boundary conditions
\begin{equation}
 \hat c(-\infty)=0,\quad
 \hat c'(-\infty)=0,\quad
 \hat c(+\infty)=c_b,\quad
 \hat c'(+\infty)=0,
\end{equation}
and continuity of $c$ and of the diffusive flux at $z=0$.

We introduce dimensionless variables
\begin{equation}
 \tilde z = z/\ell_D,\quad
 \tilde c = c/c_b,\quad
 \varepsilon = c_h/c_b,\quad
 \lambda = \ell_D/\ell_d = \ell_a/\ell_D,
\end{equation}
where $\ell_D = \sqrt{D c_b / (\gamma\rho_0)}$ and
$\ell_d = D/v$. Dropping tildes for readability, 
Eq.~\eqref{eq:tw-dimensional} becomes
\begin{equation}
 c''(z) + \lambda\, c'(z) - f\bigl(c(z)\bigr)\,\Theta(-z) = 0,
 \label{eq:tw-nondim}
\end{equation}
with $c(-\infty)=0$ and $c(+\infty)=1$. A unique value of $\lambda$
generates a trajectory connecting these two limits.

We now analyze the steady traveling wave in the sharp--interface limit
$\varepsilon \ll 1$. In this regime the Monod response is approximated
piecewise as
\begin{equation}
 f(c) \simeq
 \begin{cases}
 0, & z > 0 \quad \text{(region I)},\\[2pt]
 1, & -\zeta < z < 0 \quad \text{(region II)},\\[2pt]
 c/c_h, & z < -\zeta \quad \text{(region III)},
 \end{cases}
\end{equation}
so Eq.~\eqref{eq:tw-nondim} can be integrated separately in regions
I--III, with integration constants fixed by continuity of $c$ and of the
diffusive flux at $z=0$ and $z=-\zeta$.

The resulting concentration profile is
\begin{equation}
 \hat c(z) =
 \begin{cases}
 \displaystyle
 1 - \dfrac{C_1}{\lambda}\,e^{-\lambda z},
 & \text{(I)},\\[0.7em]
 \displaystyle
 \dfrac{\varepsilon \lambda + z + \zeta}{\lambda}
  - \dfrac{C_2}{\lambda}\bigl(e^{-\lambda z} - e^{\lambda \zeta}\bigr),
 & \text{(II)},\\[0.8em]
 \displaystyle
 \varepsilon \exp\!\left[
 \tfrac{1}{2}\Bigl(-\lambda + \Delta\Bigr)(z + \zeta)
 \right],
 & \text{(III)},
 \end{cases}
 \label{eq:c-piecewise}
\end{equation}
where we have defined
\begin{equation}
 \Delta \equiv \sqrt{\frac{4}{\varepsilon} + \lambda^{2}}.
\end{equation}
It is convenient to introduce
\begin{equation}
 B \equiv 1 + \lambda\bigl[(\varepsilon-1)\lambda + \zeta\bigr],
\end{equation}
so that the constants $C_1$ and $C_2$ can be written compactly as
\begin{equation}
  C_1 = \frac{1 - e^{-\lambda\zeta} B}{\lambda},
  \qquad
  C_2 = -\frac{e^{-\lambda\zeta} B}{\lambda}.
\end{equation}

The active-layer width $\zeta$ follows from the matching conditions and
is given by
\begin{equation}
 \zeta
 = \lambda - \frac{\varepsilon \lambda}{2}
   - \frac{1}{2}\sqrt{\varepsilon\bigl(4 + \varepsilon \lambda^{2}\bigr)}.
 \label{eq:zeta}
\end{equation}

\subsection{Pressure field in the base state}

In nondimensional form, the pressure inside the colony satisfies
\begin{equation}
 p''(z) = -f\bigl(c(z)\bigr),
\end{equation}
with $p$ continuous and $\partial_z p$ continuous across $z=-\zeta$,
and with the front speed given by
\begin{equation}
 v = -p'(0) = \lambda.
\end{equation}

In the piecewise limit, $p(z)$ can be obtained by integrating the
corresponding piecewise constant forcing. Using the same partition into
regions II ($-\zeta < z < 0$) and III ($z < -\zeta$), and choosing the
additive constant such that $p(0)=0$, one finds

\begin{equation}
 \hat P(z) =
 \begin{cases}
 \displaystyle
 -\frac{1}{2}\,z\bigl(2\lambda + z\bigr),
 & \text{(II)},\\[0.8em]
 \displaystyle
 -\frac{4\,\exp\!\left[\tfrac{1}{2}(-\lambda + \Delta)(z+\zeta)\right]}
           {(\lambda - \Delta)^{2}}
    + C_0,
 & \text{(III)},
 \end{cases}
 \label{eq:P-II}
\end{equation}
with $C_0= \tfrac{1}{2}\Bigl(  \varepsilon^{2}\lambda^{2} + \varepsilon \lambda S + \zeta(S + \zeta)+ \varepsilon(2 + \lambda \zeta)\Bigr)$

where we also introduced
\begin{equation}
 S \equiv \sqrt{\varepsilon\bigl(4 + \varepsilon \lambda^{2}\bigr)}.
\end{equation}
These expressions determine the base-state front speed $v=\lambda$ and
pressure profile $p^*(z)$.

For later reference, the height $H$ above the front where the
concentration deviates from the bulk value by a small fraction
$\delta c$ is obtained from Eq.~\eqref{eq:c-piecewise} as
\begin{equation}
 H = \frac{1}{\lambda}\,
     \ln\!\left[
        \frac{e^{\lambda^{2}}-1}{\delta c\, \lambda^{2}}
     \right].
 \label{eq:H-def}
\end{equation}

\subsection{Linear stability analysis and dispersion relation}

We now perturb the traveling--wave solution $c^*(z)$, $p^*(z)$ and the
front position $h^*$ by small lateral deformations of amplitude
$\eta \ll 1$:
\begin{align}
 c(x,z,t) &= c^*(z) + \eta\, c_{1}(z,t)\cos(kx),\\
 p(x,z,t) &= p^*(z) + \eta\, p_{1}(z,t)\cos(kx),\\
 h(x,t)   &= h^* + \eta\, h_{1}(t)\cos(kx),\\
 \zeta(x,t) &= \zeta^* + \eta\, \zeta_{1}(t)\cos(kx),
\end{align}
where $h$ is the (dimensionless) front height measured from $z=0$, and
$c^*,p^*,h^*,\zeta^*$ denote the quasi-steady base state.

Linearizing the nondimensional reaction--diffusion and Darcy equations
around this base state and retaining terms of $O(\eta)$ yields
\begin{align}
 \bigl(\partial_{z}^{2} - k^2\bigr) c_1
 &= f_1(c)\,\Theta(-z)
    - \lambda\, \partial_z c_1, \label{eq:c1-pert}\\
 \bigl(\partial_{z}^{2} - k^2\bigr) p_1
 &= \tilde{\alpha}\, f_1(c)\,\Theta(-z),
 \label{eq:p1-pert}
\end{align}
where $\tilde{\alpha}$ is the nondimensional activity parameter.

The boundary conditions on $c_1$ follow from continuity of the field and
its flux at the moving interfaces, expanded to first order in $\eta$:
\begin{equation*}
\label{eq:bc-c1}
\begin{cases}
\displaystyle \lim_{z\to+\infty} c_1(z,t) = 0,\\[4pt]
\left.c_1\right|_{0^{+}} + h_1\,\left.\partial_{z} c^*\right|_{0^{+}}
= \left.c_1\right|_{0^{-}} + h_1\,\left.\partial_{z} c^*\right|_{0^{-}},\\[4pt]
\left.\partial_{z}c_1 \right|_{0^{+}} + h_1\,\left.\partial^2_{z} c^*\right|_{0^{+}}
= \left.\partial_{z}c_1\right|_{0^{-}} + h_1\,\left.\partial^2_{z} c^*\right|_{0^{-}},\\[4pt]
\left.c_1\right|_{\zeta^{+}} = \left.c_1\right|_{\zeta^{-}},\\[4pt]
\left.\partial_{z}c_1 \right|_{\zeta^{+}} = \left.\partial_{z}c_1\right|_{\zeta^{-}},\\[4pt]
\displaystyle \lim_{z\to-\infty} c_1(z,t) = 0.
\end{cases}
\end{equation*}
Similarly, for the pressure perturbation
\begin{equation*}
\label{eq:bc-p1}
\begin{cases}
\left.p_1\right|_{z=0} = -h_1\, \partial_z p^*(0),\\[4pt]
\left.p_1\right|_{\zeta^{+}} = \left.p_1\right|_{\zeta^{-}},\\[4pt]
\left.\partial_{z} p_1\right|_{\zeta^{+}} = \left.\partial_{z} p_1\right|_{\zeta^{-}},\\[4pt]
\displaystyle \lim_{z\to-\infty} \partial_z p_1(z,t) = 0.
\end{cases}
\end{equation*}
In the sharp-interface approximation the $\zeta_1$ terms drop out
because $\partial_{z} c^*$ and $\partial^2_{z} c^*$ are continuous
across $z=\zeta$ by construction.

Expanding the growth function to first order in $\eta$ gives
\begin{equation*}
 f_1(c) = \left. \frac{df}{dc} \right|_{c=c^*} c_1 \cos(kx)
 \Rightarrow 
 f_1 =
 \begin{cases}
 0, & \text{II},\\[4pt]
 c_1 \cos(kx)/c_h, & \text{III},
 \end{cases}
\end{equation*}
so that, in the piecewise limit, only the depleted interior
(region III) responds linearly to concentration fluctuations.

For compactness we introduce the $k$–dependent quantities
$\alpha = \sqrt{1+4k^2}$, $\chi   = \sqrt{\varepsilon+4\varepsilon k^2}$, $\Sigma = \sqrt{4+\varepsilon+4\varepsilon k^2}$, $\Lambda = \sqrt{1+\tfrac{4}{\varepsilon}+4k^2}$, $\Phi    = \sqrt{\varepsilon\bigl(4+\varepsilon+4\varepsilon k^2\bigr)}$, $Q   = 2 + \varepsilon \alpha^{2} - \chi\,\Sigma$ and $H_0 = \varepsilon\lambda + \sqrt{\varepsilon\bigl(4+\varepsilon\lambda^{2}\bigr)}+ 2\zeta.$

In each region, Eqs.~\eqref{eq:c1-pert}--\eqref{eq:p1-pert} admit
exponential solutions. It is convenient to introduce the $k$--dependent
decay rates
\begin{align*}
 &\mu(k) = \tfrac{1}{2}\bigl(1 + \sqrt{1+4k^2}\bigr), \\
 &\nu(k) = \tfrac{1}{2}\left(-1 + \frac{\sqrt{4+\varepsilon + 4\varepsilon k^2}}{\sqrt{\varepsilon}}\right),
\end{align*}
so that the concentration perturbation can be written as
\begin{equation*}
  c_1(z) = h_1
  \begin{cases}
  \displaystyle
  \frac{1}{2\alpha}\,
  e^{-\frac{1+\alpha}{2}\,z - \alpha\zeta}
  \bigl(Q - 2 e^{\alpha\zeta}\bigr),\quad
   z>0, \\[0.8em]
  \displaystyle
  \frac{1}{2\alpha}\,
  e^{-\frac{1+\alpha}{2}\,z - \alpha\zeta}
  \bigl(Q - 2 e^{\alpha(z+\zeta)}\bigr),\quad
   -\zeta<z<0,\\[0.8em]
  \displaystyle
  \frac{\sqrt{\varepsilon}}{2}\,(\chi-\Sigma)\,
  \exp\!\left[
    \frac{1}{2}\Bigl(
      -z - \alpha\zeta + \Lambda(z+\zeta)
    \Bigr)
  \right],
   z<-\zeta
  \end{cases}
  \label{eq:c1-piecewise}
\end{equation*}

Inside the colony ($z\le 0$) the pressure perturbation is
\begin{equation*}
  p_1(z) =
  \begin{cases}
  \displaystyle
  \frac{1}{2} e^{-k z}
  \Bigl[
    2 \beta_2\bigl(-1+e^{2kz}\bigr) + h_1 H_0
  \Bigr],
  \quad -\zeta<z<0,\\[0.8em]
  \displaystyle
  \beta_3 e^{kz}
  - \frac{2 \beta_1}{2+\varepsilon-\Phi}\,
    \exp\!\left[
      \frac{1}{2}\bigl(-1+\Lambda\bigr) z
    \right],
   z<-\zeta,
  \end{cases}
  \label{eq:p1-piecewise}
\end{equation*}
with coefficients $\beta_1 = \tfrac{\sqrt{\varepsilon}\,h_1}{2}(\chi-\Sigma)
       \exp\!\bigl[(-\chi+\Sigma)\zeta/(2\sqrt{\varepsilon})\bigr]$,
$\beta_2 = \tfrac{1}{2}\Bigl[
       \dfrac{\beta_1 e^{-\frac{1}{2}(-1+2k+\Lambda)\zeta}
              \bigl(\sqrt{\varepsilon}(1+2k)-\Sigma\bigr)}
              {\sqrt{\varepsilon}\,k\,(2+\varepsilon-\Phi)}
       + h_1 H_0\Bigr]$,
and
$\beta_3 = \beta_2(1-e^{2k\zeta})
       + \dfrac{2\beta_1 e^{(\tfrac{1}{2}+k-\Lambda/2)\zeta}}
               {2+\varepsilon-\Phi}
       + \tfrac{1}{2}e^{2k\zeta}h_1 H_0$.

\subsection{Front kinematics and dispersion relation}

The front dynamics follow from the kinematic condition
\begin{equation*}
 \dot{h} = -\left.\partial_z p\right|_{z=0},
\end{equation*}
which we expand to linear order in $\eta$ by evaluating the pressure
gradient at $z = \varepsilon h_1 \cos(kx)$. This yields
\begin{align*}
 &\dot{h}_0 + \dot{h}_1\cos(kx)=\\
 &= -\left[p_0'(0) + p_0''(0)\,\varepsilon h_1\cos(kx)
        + p_1'(0)\cos(kx)\right],\\
 \Rightarrow\quad
 \dot{h}_1
 &= -\bigl[h_1\, p_0''(0) + p_1'(0)\bigr],
\end{align*}
where $p_0(z) = p^*(z)$ is the base-state pressure.
Using Eq.~\eqref{eq:P-II} one finds $p_0''(0)=-1$, so that
\begin{equation*}
 \dot{h}_1 = \omega(k)\, h_1,
 \qquad
 \omega(k) = 1 - p_1'(0)/h_1.
\end{equation*}

Evaluating $p_1'(0)$ from the piecewise solutions and applying the
matching conditions gives a closed-form dispersion relation. To keep the
expression compact, we define $\theta(k) = \tfrac{1}{4}\bigl[(-1+2k+\Lambda)\bigl((\varepsilon-2)\lambda+S\bigr)- 2\alpha\,\zeta + (2\Sigma/\sqrt{\varepsilon})\,\zeta\bigr]$.

With these definitions, the growth rate can be written in the compact form
\begin{equation}
 \omega(k)
 = 1 - k\lambda
   - \frac{e^{\theta(k)}\, (\Sigma(k) - \sqrt{\varepsilon}\,(1+2k))\,(\Sigma-\chi)}{4+2\varepsilon-2\Phi}
 \label{eq:omega-compact}
\end{equation}
The fastest-growing wavenumber $k_{\mathrm{max}}$ maximizes
$\omega(k)$, and the corresponding wavelength is
\begin{equation}
 \lambda_{\mathrm{max}} = \frac{2\pi}{k_{\mathrm{max}}}.
\end{equation}
In a one-dimensional strip of length $L$, the number of whole
wavelengths that can fit is
\begin{equation}
 N = \left\lfloor \frac{L}{\lambda_{\mathrm{max}}} \right\rfloor
   = \left\lfloor \frac{k_{\mathrm{max}}L}{2\pi} \right\rfloor,
\end{equation}
which sets the typical number of fingers selected by the linear
instability.

The dispersion relation is plot for different values of $\ell$ where for the velocity $v$ we used the one measured from the simulations \cref{pan:Velocity-front} averaged through different values of $\mu$ for each $\ell$. Here increasing $c_h$ makes growth more homogeneous and suppresses the protrusions.

\clearpage

\end{document}